\documentclass[referee,pdflatex,sn-aps]{sn-jnl}

\usepackage{graphicx}%
\usepackage{multirow}%
\usepackage{amsmath,amssymb,amsfonts}%
\usepackage{amsthm}%
\usepackage{mathrsfs}%
\usepackage[title]{appendix}%
\usepackage{textcomp}%
\usepackage{manyfoot}%
\usepackage{booktabs}%
\usepackage{algorithm}%
\usepackage{algorithmicx}%
\usepackage{algpseudocode}%
\usepackage{listings}%

\usepackage[utf8]{inputenc}
\usepackage[caption=false]{subfig}
\usepackage[dvipsnames]{xcolor}
\usepackage{bbm}
\usepackage{slashed}
\usepackage{physics}
\usepackage{xspace}
\usepackage{orcidlink}
\usepackage{comment}
\usepackage{glossaries}
\setacronymstyle{long-short}
\glsdisablehyper

\usepackage{hyperref}
\hypersetup{colorlinks=true,
    linkcolor=magenta,
    citecolor=blue,
    urlcolor=cyan,
    pdfpagemode=FullScreen,}

\newacronym{pionless}{EFT$_\slashed \pi$}{pionless effective field theory}
\newacronym{eft}{EFT}{effective field theory}
\newacronym{qed}{QED}{quantum electrodynamics}
\newacronym{qcd}{QCD}{quantum chromodynamics}
\newacronym{nrqed}{NRQED}{nonrelativistic QED}
\newacronym{vnrqed}{vNRQED}{velocity NRQED}
\newacronym{nrqcd}{NRQCD}{nonrelativistic QCD}
\newacronym{ChiPT}{\ensuremath{\chi}PT}{chiral perturbation theory}
\newacronym{ChiEFT}{\ensuremath{\chi}EFT}{chiral effective field theory}
\newacronym{BSM}{BSM}{Beyond the Standard Model}
\newacronym{NN}{\ensuremath{N\!N}\xspace}{nucleon-nucleon}
\newacronym{vRG}{vRG}{velocity renormalization group}
\newacronym{lo}{LO}{leading order}
\newacronym{nlo}{NLO}{next-to-leading order}
\newacronym{n2lo}{\ensuremath{\text{N}^2\text{LO}}}{next-to-next-to-leading order}
\newacronym{n3lo}{\ensuremath{\text{N}^3\text{LO}}}{next-to-next-to-next-to-leading order}
\newacronym{n4lo}{\ensuremath{\text{N}^4\text{LO}}}{next-to-next-to-next-to-next-to-leading order}
\newacronym{dimreg}{DimReg}{dimensional regularization}
\newacronym{lec}{LEC}{low energy coefficient}
\newacronym{av18}{AV18}{Argonne \ensuremath{v18}}

\newcommand{\oneS}{{{}^{1}\!S_0}}
\newcommand{\threeS}{{{}^{3}\!S_1}}

\newcommand{\del}{\nabla}

\newcommand{\calL}{\ensuremath{\mathcal{L}}}

\newcommand{\1}{\mathbbm{1}}

\newcommand{\NN}{\ensuremath{N\!N}\xspace}

\begin{document}

\title[Radiative corrections and the renormalization group for the two-nucleon interaction in effective field theory]{Radiative corrections and the renormalization group for the two-nucleon interaction in effective field theory}

\author*[1]{\fnm{Thomas R.} \sur{Richardson\,\orcidlink{0000-0001-6314-7518}}}\email{richardt@uni-mainz.de}

\author[1]{\fnm{Immo C.} \sur{Reis\,\orcidlink{0009-0008-3566-6095}}}\email{imreis@students.uni-mainz.de}
\equalcont{These authors contributed equally to this work.}

\affil[1]{\orgdiv{Institut f\"ur Kernphysik and PRISMA$^+$ Cluster of Excellence}, \orgname{Johannes Gutenberg-Universit\"at}, \orgaddress{\postcode{55128}, \city{Mainz}, \country{Germany}}}


\abstract{We use a combination of effective field theory and the renormalization group to determine the impact of radiative corrections on the nucleon-nucleon potential and the binding energy of the deuteron.
In order to do so, we present a modified version of pionless effective field theory inspired by earlier work in nonrelativistic quantum electrodynamics.
The renormalization group improvement of the deuteron binding energy leads to a shift on the order of a few percent and is consistent with the experimental value.
This work serves as a starting point for a dedicated study of radiative corrections in few-body systems relevant for precision tests of the Standard Model in an effective field theory framework.}


\maketitle


\section{Introduction}
Modern experiments that rely on few-nucleon systems such as $\beta$-decay \cite{pocanicNabMeasurementPrinciples2009,dubbersCleanBrightVersatile2008}, $\mu$-capture \cite{andreevMuonCaptureDeuteron2010,kammelMuSunMuonCapture2021}, and muonic atom spectroscopy \cite{antogniniMuonicAtomSpectroscopyImpact2022,antogniniProtonStructureMeasurement2013,pohlLaserSpectroscopyMuonic2016,pohlSizeProton2010,krauthMeasuringAparticleCharge2021,thecremacollaborationHelionChargeRadius2023} are reaching subpercent-level precision. 
Thus, these experiments can provide stringent tests for the Standard Model in low energy systems and possibly shed light on new physics.
However, a correct interpretation of the experimental results requires a thorough theoretical understanding and delineation of the different effects involved.

In particular, these experiments are sensitive to radiative corrections from electrodynamics.
In the context of muonic atom spectroscopy, a subset of these effects has been the subject of significant theoretical interest \cite{antogniniProtonStructureOut2022,jiInitioCalculationNuclear2018,pachuckiComprehensiveTheoryLamb2023}.
It is customary to include radiative corrections through finite nuclear size effects and the exchange of two or more photons between the nucleus and the bound muon.
The nuclear wavefunctions and currents, however, only include electromagnetic effects implicitly by fitting the parameters of the nuclear Hamiltonian and currents to data.
Because of this, there is no way to distill how much of an observable comes from \gls{qcd} as opposed to electroweak interactions.

In the case of $\beta$-decays, this topic has received renewed interest in recent years with respect to single-neutron $\beta$-decay \cite{andoNeutronBetaDecay2004,ciriglianoPioninducedRadiativeCorrections2022,sengDispersiveEvaluationInner2019,sengReducedHadronicUncertainty2018,ciriglianoEffectiveFieldTheory2023,czarneckiRadiativeCorrectionsNeutron2019,hayenStandardModelMathcal2021,gorchteinDispersionRelationAnalysis2021}.
Interestingly, Ref.~\cite{ciriglianoPioninducedRadiativeCorrections2022} finds a percent level shift in the nucleon axial coupling $g_A$ due to radiative corrections that shifts the lattice \gls{qcd} determination of $g_A$ closer to the more precise experimental value.
This represents a significant step towards disentangling the myriad of effects involved in neutron $\beta$ decay in terms of Standard Model parameters.

The goal of this work is to begin bridging the gap in few-nucleon systems with \gls{eft} techniques.
We use a combination of \gls{pionless} \cite{kaplanNucleonNucleonScattering1996, kaplanNewExpansionNucleonnucleon1998, kaplanTwoNucleonSystems1998, vankolckEffectiveFieldTheory1999b, chenNucleonnucleonEffectiveField1999, beaneHadronsNucleiCrossing2001a,rupakPrecisionCalculationNp2000,bedaqueEffectiveFieldTheory2002, hammerNuclearEffectiveField2020a} and the \gls{vRG} \cite{lukeRenormalizationGroupScaling2000} developed for \gls{nrqed} \cite{caswellEffectiveLagrangiansBound1986}.
This theory is valid for momenta $p \ll m_\pi$, where $m_\pi$ is the pion mass, which is in the regime relevant for many of these experiments.
Certain aspects of this work can also be applied in chiral \gls{eft} \cite{hammerNuclearEffectiveField2020a,epelbaumModernTheoryNuclear2009a,epelbaumHighprecisionNuclearForces2019,epelbaumSemilocalNuclearForces2022,machleidtChiralEffectiveField2011a}, which has a larger radius of convergence. 
On the other hand, the entire framework can immediately be applied in an \gls{eft} for halo nuclei \cite{hammerNuclearEffectiveField2020a, bertulaniEffectiveFieldTheory2002, hammerTheoryHaloNuclei2022, hammerEffectiveFieldTheory2017a} with trivial modifications.

In this work, we use the renormalization group to sum the leading logarithm series in $\alpha$, where $\alpha = e^2/4 \pi$ is the fine structure constant, into the coefficients of the neutron-proton potential.
The running couplings that follow from the \gls{vRG} equations can in principle be embedded in \textit{ab initio} calculations using few- or many-body methods.
The electromagnetic renormalization of this potential will generate isospin breaking contributions in the electroweak processes already described (see, e.g., Refs.~\cite{sengItItInitio2023,townerImprovedCalculationIsospinsymmetrybreaking2008,millerIsospinsymmetrybreakingCorrectionsSuperallowed2008, gorchteinSuperallowedNuclearBeta2023} in the context of nuclear $\beta$-decays).

To illustrate the impact of the running induced by the radiative corrections, we use renormalization group improved perturbation theory to calculate the deuteron binding energy and compare the result to the fixed order calculation.
In order to generate numerical results, the \gls{vRG} equations require a boundary condition to fix the final value of low energy coefficients (LECs).
Ideally, the LECs in a nuclear \gls{eft} in the absence of electroweak effects would be determined by lattice \gls{qcd} rather than data.
However, available few-nucleon lattice calculations have greater than physical $m_\pi$ and the uncertainties are quite large.
In the meantime, we make use of the scattering parameters of the phenomenological \gls{av18} potential without electromagnetic interactions found in Table VIII of Ref.~\cite{wiringaAccurateNucleonnucleonPotential1995} (also see Ref.~\cite{tuminoCoulombfree1S0Scattering2023}).
Here, we find that \gls{vRG} improvement drives a percent level shift in the deuteron binding energy.
This observation is consistent with the \gls{av18} potential, but it recasts the impact of electromagnetic corrections to the nucleon-nucleon ($\NN$) interactions in terms of a modern \gls{eft} with the full machinery of the renormalization group.


\section{Reorganizing pionless effective field theory}
Now, we recast \gls{pionless} in the language of \gls{vnrqed} \cite{lukeRenormalizationGroupScaling2000}.
In \gls{pionless} it is typical to count powers of the momentum $p$, but in \gls{nrqed} powers of velocity $v = p/M_N$, where $M_N$ is the nucleon mass, are counted.
However, some care is needed in judging the scaling of electromagnetic interactions relative to the strong \NN interaction; we revisit this at the end of this section.
The relevant (energy, momentum) scales are then expressed as hard $(m_\pi, m_\pi)$, soft $(M_N v, M_N v)$, ultrasoft $(M_N v^2, M_N v^2)$, and potential $(M_N v^2, M_N v)$.
Power counting issues are avoided by splitting the photon into multiple modes describing the soft and ultrasoft regions and multipole expanding the ultrasoft modes \cite{labelleEffectiveFieldTheories1998a,grinsteinEffectiveFieldTheory1998,lukePowerCountingDimensionally1998,lukeBoundStatesPower1997,griesshammerThresholdExpansionDimensionally1998,griesshammerPowerCountingBeta2000}.
The potential photons can be integrated out because they are far off-shell; their effects are encoded in the coefficients of four-nucleon operators.

After performing a nonrelativistic reduction, the four-momentum of the nucleon is decomposed as
    \begin{equation}
            \label{eq:background:nrqed:nucleon_momentum}
        P = (0, \vb p) + (k_0, \vb k) \, ,
    \end{equation}
where $\vb p \sim M_N v$ is the soft component of the momentum and $k \sim M_N v^2$ is the residual four-momentum on the ultrasoft scale.
The on-shell condition becomes $k_0 = \vb p^2/2 M_N$.
The nucleon field is now written as $N_{\vb p} (x)$ where $\vb p$ is a soft label, $x$ is the Fourier conjugate of the residual momentum $k$, and $N$ is an isodoublet of the proton and neutron.

The photon field is also split into a soft field $A^\mu_p (k)$ with soft label four-momentum $p$ and a residual four-momentum $k$ and an ultrasoft field $A^\mu$.
Conservation of energy excludes interactions of the type $A_q N_{\vb p}^\dagger N_{\vb p}$, i.e., only vertices with two soft photon lines are allowed.
The kinetic term of the photon field is split into 
    \begin{align}
        \calL \supset & - \frac{1}{4} F_{\mu \nu} F^{\mu \nu} + \sum_p \abs{p^\mu A_p^\nu - p^\nu A_p^\mu}^2 \, ,
    \end{align}
where the field strength tensor $F_{\mu \nu}$ contains only ultrasoft photons.

Reparameterization invariance implies that derivatives acting on the nucleon fields appear in the combination $i \vb p + \vb D$, where $\vb p$ acts on the soft label and $\vb D$ is a covariant derivative acting on the residual piece of the nucleon field.
In the kinetic term for the nucleon, the term $\left( \vb p - i \vb D \right)^2$ should be expanded, which is equivalent to the multipole expansion, and only the $\vb p^2$ should be kept in the leading order propagator.
Therefore, the nucleon propagator will be
    \begin{equation}
        S(k_0, \vb p) = \frac{i}{k_0 - \frac{\vb p^2}{2 M_N} + i \epsilon} \, .
    \end{equation}
Terms containing factors of $\vb p \cdot \del$ or $\del^2$ are treated as perturbations.

While \gls{pionless} is usually formulated in an isospin basis, we find it more convenient to study the ultrasoft renormalization of the potential in terms of physical neutron and proton fields $n$ and $p$, respectively.
The LECs can of course be translated into the isospin basis after the renormalization has been carried out.
In this basis, the proton-neutron potential is written as
    \begin{align}
            \label{eq:potential}
        V_{pn} & = \sum_{v=-1} \sum_{\vb p', \vb p} V^{(v, pn)}_{abcd} (\vb p', \vb p) p^\dagger_{\vb p', a} p_{\vb p, b} n^\dagger_{- \vb p', c} n_{-\vb p, d} \, ,
    \end{align}
where $v$ tracks the order in the velocity expansion of each coefficient, and $a$, $b$, $c$, and $d$ are spin indices for the neutron and proton fields.
The \gls{lo}, \gls{nlo}, and \gls{n2lo} potential coefficients in the S-wave are given by
    \begin{align}
        V^{(-1, pn)}_{abcd} & = C_{0, pn}^{(\threeS)} P^{(1)}_{ab, cd} + C_{0, pn}^{(\oneS)} P^{(0)}_{ab, cd} \, , \label{eq:LOpotential}\\
        V^{(0, pn)}_{abcd} & = \frac{1}{2} \left( \vb p'^2 + \vb p^2 \right) \left[ C_{2, pn}^{(\threeS)} P^{(1)}_{ab, cd} + C_{2, pn}^{(\oneS)} P^{(0)}_{ab, cd} \right] \, , \label{eq:NLOpotential}\\
        V^{(1, pn)}_{abcd} & = \frac{1}{4} \left( \vb p'^2 + \vb p^2 \right)^2 \left[ C_{4, pn}^{(\threeS)} P^{(1)}_{ai, bj} + C_{4, pn}^{(\oneS)} P^{(0)}_{ab, cj} \right] \label{eq:N2LOpotential} \, ,
    \end{align}
where the projection operators are given by
    \begin{align}
        P^{(1)}_{ab, cd} & = \frac{1}{4} \left( 3 \delta_{ab} \delta_{cd} + \sigma^i_{ab} \sigma^i_{cd} \right) \, , \\
        P^{(0)}_{ab, cd} & = \frac{1}{4} \left( \delta_{ab} \delta_{cd} - \sigma^i_{ab} \sigma^i_{cd} \right) \, .
    \end{align}
Note that our definition of $C_4$ is a linear combination of $C_4 + \tilde C_4$ that appears in the literature (see for example Refs.~\cite{chenNucleonnucleonEffectiveField1999,beaneHadronsNucleiCrossing2001a,rupakPrecisionCalculationNp2000}).
The $V^{(0)}$ potential should also be supplemented with a correction to the Coulomb potential that arises from a potential photon coupled to the proton charge and the neutron charge radius as well as a term that couples the neutron and proton magnetic moments; however, these terms are also suppressed by a factor of $\alpha$ bringing the overall sizes to $O(\alpha v^0)$.
Therefore, the velocity power counting suggests that these contributions are higher order than what we will consider.

The neutron-neutron potentials have an identical structure with respect to the purely strong interactions.
The part of the potential that arises from potential photon exchange is $O(\alpha v^2)$.
The strong part of the proton-proton potential is also identical to the proton-neutron potential. 
However, we have to add the Coulomb potential to the leading order term.
    \begin{align}
        V^{(-1, pp)}_{abcd} & \supset \sum_{\vb p', \vb p} \frac{4 \pi \alpha}{(\vb p' - \vb p)^2} \delta_{ab} \delta_{cd}.
    \end{align}
In the remainder of this work we will only consider the proton-neutron channel, so we only retain $V_{pn}$ from Eq.~\eqref{eq:potential}.
All together, the Lagrangian we will work with is
    \begin{align}
            \label{eq:nrqed_pionless:lagrangian}
        \calL = & \sum_{\vb p} N_{\vb p}^\dagger \left( i D_0 - \frac{\left( \vb p -i \vb D \right)^2}{2 M_N} \right) N_{\vb p} - \frac{1}{4} F_{\mu \nu} F^{\mu \nu} + \sum_{\vb p} \abs{p^\mu A_p^\nu - p^\nu A_p^\mu}^2 - \color{blue}{V_{pn}} \nonumber \\
        & - \frac{4 \pi \alpha}{2 M_N} \sum_{q, q', \vb p, \vb p'} \vb A_{q'} \cdot \vb A_{q} N^\dagger_{\vb p'} Q N_{\vb p} + \frac{e}{2 M_N} \epsilon^{ijk} \left( \del^j A^k \right) \sum_{\vb p} N_{\vb p}^\dagger \sigma^i \left[ \kappa_0 + \kappa_1 \tau^3 \right] N_{\vb p} \, ,
    \end{align}
where we have reverted to the nucleon isodoublet $N_{\vb p}$ in the single-nucleon terms to condense the notation, $Q = (\1 + \tau_3)/2$ is the nucleon charge matrix, and $V_{pn}$ is the neutron-proton potential defined in terms of neutron and proton fields in Eq.~\eqref{eq:potential} with coefficients from Eqs.~\eqref{eq:LOpotential}-\eqref{eq:N2LOpotential}.

\begin{figure}
    \centering
    \subfloat[]{%
    \includegraphics[width=0.9\textwidth]{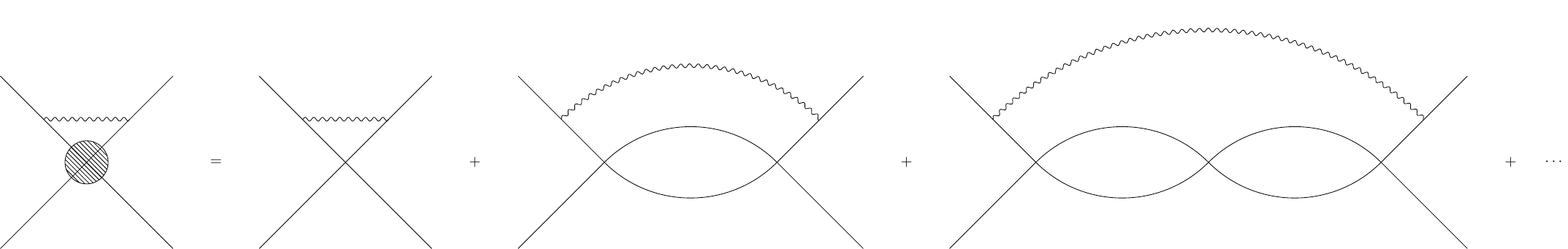}%
}
\caption{$O(\alpha/v)$ diagrams that contribute to the anomalous dimension of the potential.}
\label{fig:diagrams}
\end{figure}
Counting powers of velocity in Feynman diagrams is fairly straightforward. 
Nucleon and soft photon propagators count as $1/v^2$ while ultrasoft photon propagators count as $1/v^4$.
The purely \NN potentials follow the standard power counting of \gls{pionless} where $Q \sim M_N v$ \cite{kaplanNewExpansionNucleonnucleon1998, kaplanTwoNucleonSystems1998}.
Finally, a soft loop has an integration measure that scales as $v^4$, a potential loop scales as $v^5$, and an ultrasoft loop scales as $v^8$.

In order to implement the \gls{vRG}, we determine the $O(\alpha/v)$ counterterms and obtain the soft and ultrasoft anomalous dimensions from \cite{lukeRenormalizationGroupScaling2000,manoharRenormalizationGroupCorrelated2000}
    \begin{align}
        \mu_U \frac{d V}{d \mu_U} & = \gamma_U \, , \\
        \mu_S \frac{d V}{d \mu_S} & = \gamma_S \, ,
    \end{align}
where $\mu_S$ is the scale introduced in dimensional regularization for the potentials and soft interactions and $\mu_U$ is the scale introduced for the ultrasoft interactions.
Through these scales we introduce the subtaction velocity $\nu$ as $\mu_S = M_N \nu$ and $\mu_U = M_N \nu^2$ so that the \gls{vRG} equation is
    \begin{align}
        \nu \frac{d V}{d \nu} & = \gamma_S + 2 \gamma_U \, .
    \end{align}
In \gls{nrqed}, this procedure is fairly easy because the fine structure constant $\alpha$ does not run and the \gls{lo} Coulomb potential is not renormalized \cite{manoharLogarithmsEnsuremathAlpha2000}.
Moreover, the $\alpha$ and $v$ expansions are identical since the average velocity in a Coulomb bound state is $O(\alpha)$.
As we will see below, because the $\alpha$ and $v$ expansions are not strictly linked in the nuclear \gls{eft} and because $\alpha$ runs due to the fact that the electron mass is not integrated out of the theory, there is a much richer structure that arises from the \gls{vRG}.

Before proceeding to the calculation of the anomalous dimensions, we comment further on the power counting in this EFT.
As previously mentioned, the $v$ and $\alpha$ expansions are not identical as they are in pure \gls{nrqed}.
However, we can estimate the relative importance of $O(\alpha)$ corrections to the strong sector.
The typical relative momentum in the deuteron is $p \sim 50$ MeV, so that the the expansion in the purely strong sector is in powers of $Q = p/m_{\pi}$ which is roughly $1/3$.
For $\alpha \sim 1/137$, a conservative estimate in this regime suggests $\alpha \sim Q^4$.
Thus, through \gls{n2lo} in the potential sector, we can safely neglect the finite contributions from the diagrams when we calculate the binding energy of the deuteron below and encode the effects from radiative correction in the running couplings via the solutions of the \gls{vRG} equations.
We expect that these contributions should be included explicitly around \gls{n3lo}.
Indeed, these contributions are already probed through the variation of the subtraction velocity.
For other processes such as proton-proton fusion where the typical momentum or velocity is somewhat smaller, the explicit radiative corrections will enter at even lower orders.

In the remainder of this work, we will focus mainly on the neutron-proton sector at $O(\alpha/v)$.
The neutron-neutron potential will be renormalized at higher orders in the $v$ expansion.
Renormalizing the proton-proton potential is much more involved.
The Coulomb interaction will generate a nonzero soft anomalous dimension for $C_0$ \cite{kongCoulombEffectsLow2000a} leading to a faster running.
Thus, we expect the \gls{vRG} to lead to interesting results in this channel.


\section{Renormalization}
The renormalization procedure in this theory is reminiscent of the role of radiation pions in \gls{eft} \cite{mehenRadiationPionsTwonucleon2000}.
However, there are several important differences.
First, we can treat both ultraviolet and infrared divergences in dimensional regularization, which simplifies the loop integrals.
Second, the neutron has no coupling to $A_0$ photons at the order we are working.

With this set-up, the basic topologies that renormalize the potential are shown in Fig.~\ref{fig:diagrams}.
In Feynman gauge, the dominant contribution, which is $O(\alpha/v)$, comes from an $A_0$ photon coupled to the proton on both the incoming and outgoing lines with insertions of the $C_0$ potential.
Inside the ultrasoft loop, an arbitrary number of \NN bubbles with only $C_0$ vertices will contribute at the same order; therefore, the internal bubble diagrams must be summed to all orders.

This infinite sum of diagrams often makes explicit renormalization of the series intractable.
The argument in the case of radiation pions is that the bubble sum should be performed before the ultrasoft integration \cite{mehenRadiationPionsTwonucleon2000}.
However, it should really be understood that the \textit{finite} parts of the bubbles are being summed, i.e., all divergences are canceled by the appropriate counterterms and the remainder is resummed.
In this case, we can actually perform this renormalization to all orders in $C_0$.

In the bubble series, each graph is divergent.
However, each graph with an odd number of \NN bubbles is ultraviolet finite and the divergence is purely infrared.
Each graph with an even number of \NN bubbles has both ultraviolet and infrared divergences which must be separated.
Specifically, a graph with $l = 2j$ bubbles, where $j$ is an integer, requires a counterterm that renormalizes the $2j$-derivative potential.
For example, the diagram with 0 \NN bubbles renormalizes the $V^{(-1)}$ potential while the diagram with 2 \NN bubbles renormalizes the $V^{(0)}$ potential.
For arbitrary $j$, the appropriate counterterm in minimal subtraction is
    \begin{align}
        \delta C_{2j} & = \frac{\alpha C_0}{2 \pi} \left( \frac{ i M_N C_0}{4 \pi} \right)^{2j} \frac{1}{j+1} \frac{1}{\epsilon} \, .
    \end{align}
For the \gls{lo} potential, we find $\gamma_{S,0} = 0$ while
    \begin{align}
            \label{eq:ultrasoft:c0}
        \gamma_{U, 0} & = \frac{1}{2 \pi} \alpha(M_N\nu^2) C_0 \, ,
    \end{align}
which leads to the \gls{vRG} equation
    \begin{align}
        \nu \frac{d C_0}{d \nu} & = \frac{1}{\pi} \alpha(M_N\nu^2) C_0.
    \end{align}
For $j \geq 1$, we find
    \begin{align}
        \gamma_{S, 2j} & \supset \frac{\alpha}{\pi} \frac{2j}{j + 1} C_0 \left( \frac{ i M_N C_0 }{4 \pi} \right)^{2j} \, , \\
        \gamma_{U, 2j} & \supset \frac{\alpha}{2 \pi} \frac{3 + 4j}{j+1} C_0 \left( \frac{ i M_N C_0 }{4 \pi} \right)^{2j} \, .
    \end{align}
There is also a contribution to the ultrasoft anomalous dimension of the $2j$-derivative operator from an insertion of the operator itself into the one-loop diagram, i.e., the first diagram on the right hand side of Fig.~\ref{fig:diagrams}.
This contribution is identical to that for $C_0$ though only with $C_{2j}$ appearing instead.
Dressing the potential vertex with additional $C_0$ interactions leads to diagrams of the same order in $v$, which will also generate contributions to the soft anomalous dimension of higher-derivative operators.
However, these contributions are still be suppressed relative to the anomalous dimensions presented here.
Retaining only the leading contribution to the anomalous dimension leads to the \gls{vRG} equation
    \begin{align}
        \nu \frac{d C_{2j}}{d \nu} & = \frac{\alpha}{\pi} \frac{3 + 6j}{j+1} C_0 \left( \frac{ i M_N C_0 }{4 \pi} \right)^{2j}.
    \end{align}

\begin{figure}
    \centering
    \includegraphics[width=0.75\textwidth]{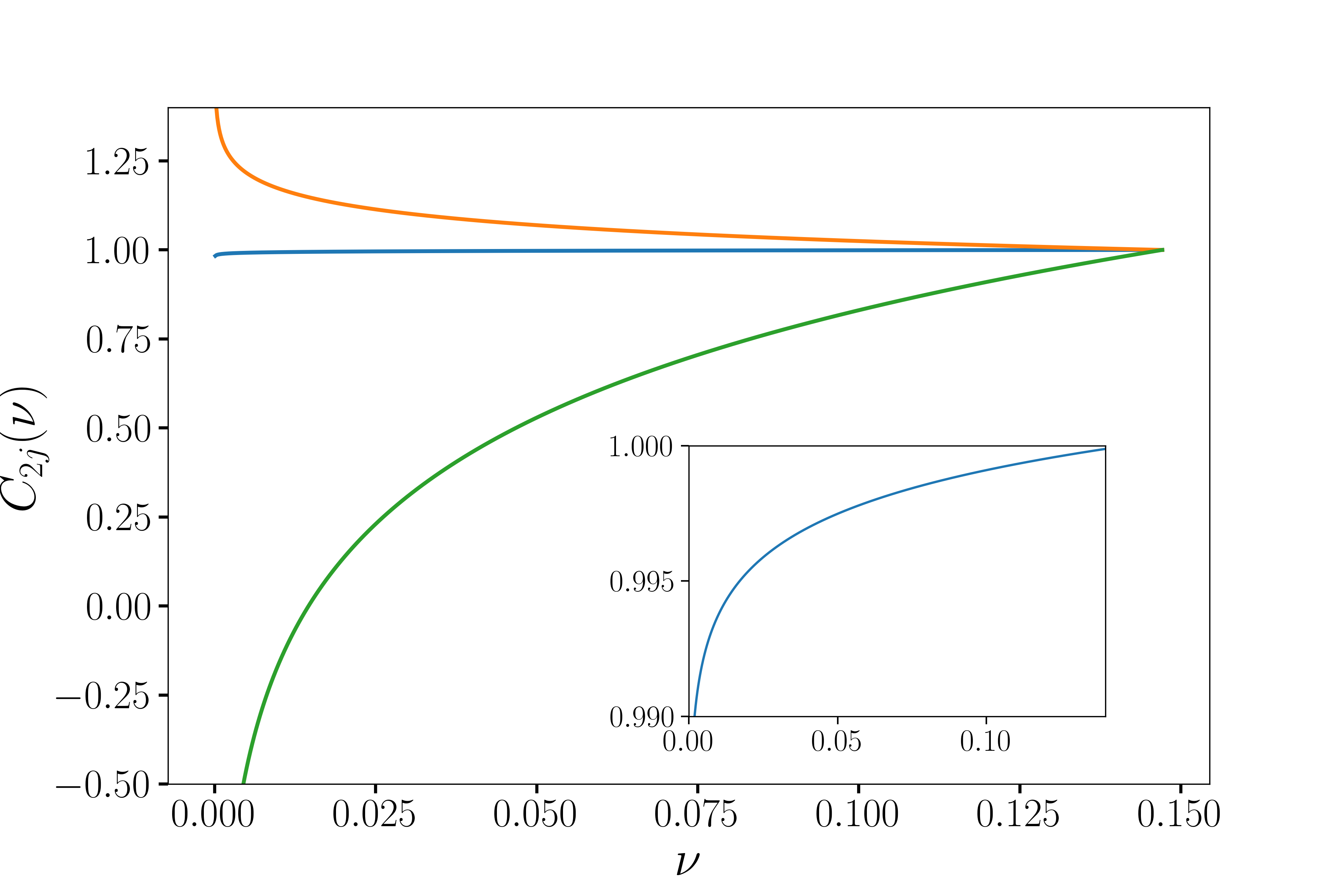}
    \caption{The running of the potential coefficients. The blue line is the running of $\hat C_0$, the orange line is the running of $\hat C_2$, and the green line is the running of $\hat C_4$.}
    \label{fig:potential_running}
\end{figure}

Integrating the \gls{vRG} equation gives
    \begin{align}
        C_0(\nu) & = C_0\left( \frac{m_\pi}{M_N} \right) \left( \frac{\alpha(M_N \nu^2)}{\alpha(m_\pi^2/M_N)} \right)^{3/4} \, , \\
        C_2(\nu) & = C_2\left( \frac{m_\pi}{M_N} \right)  - \frac{3}{2} \left( \frac{M_N}{4 \pi} \right)^2 C_0^3\left( \frac{m_\pi}{M_N} \right) \left[ \left( \frac{\alpha(M_N \nu^2)}{\alpha(m_\pi^2/M_N)} \right)^{9/4} - 1 \right] \, , \\
        C_4(\nu) & = C_4\left( \frac{m_\pi}{M_N} \right) + \left( \frac{M_N}{4 \pi} \right)^4 C_0^5\left( \frac{m_\pi}{M_N} \right) \left[ \left( \frac{\alpha(M_N \nu^2)}{\alpha(m_\pi^2/M_N)} \right)^{15/4} - 1 \right] \, .
    \end{align}
In Fig.~\ref{fig:potential_running}, we show the running of the potential LECs normalized as
    \begin{align}
        \hat C_{2j}( \nu ) & = \frac{C_{2j} (\nu)}{C_{2j} (m_\pi/M_N)} \, ,
    \end{align}
where the normalization condition is discussed below in Eqs.~\eqref{eq:scattering_parameter:a} through Eq.~\eqref{eq:normalization:c4}.
The zero-derivative potential runs very slowly while $\hat C_2$ differs by several percent from its value at the hard scale when $\nu < 0.06$.
The running of $\hat C_4$ is significantly faster; it changes by nearly $50\%$ when $\nu \sim 0.06$


\section{Impact in the deuteron}
The two-point correlation function for the deuteron is given by \cite{kaplanPerturbativeCalculationElectromagnetic1999a}
    \begin{align}
        G(\bar E) & = \frac{ \Sigma(\bar E) }{1 + i C_0 \Sigma(\bar E)} \, ,
    \end{align}
where $\Sigma$ is the self-energy of the deuteron and consists of irreducible diagrams in the sense that they do not fall apart when cut at a $C_0$ vertex.
The self-energy is expanded as
    \begin{align}
        \Sigma(\bar E) & = \Sigma_1(\bar E) + \Sigma_2(\bar E) + \Sigma_3(\bar E) + \cdots
    \end{align}
where $\bar E$ is the center-of-mass energy and the dots stand for higher order terms.
The subscript indicates the order of the contribution in the velocity expansion.

On one hand, the two-point function has the form
    \begin{align}
        G(\bar E) & = \frac{i \tilde Z}{\bar E + \tilde B} \, ,
    \end{align}
where $\tilde Z$ and $\tilde B$ are the wave-function renormalization and binding energy in the full theory, respectively.
The perturbative corrections lead to shifts in the wave function renormalization $\tilde Z = Z + \delta Z$ and the binding energy $B = B + \delta B$, where $Z$ and $B$ are the \gls{lo} results for the wave-function renormalization and binding energy, respectively.
Expanding the two-point function leads to
    \begin{align}
            \label{eq:2pt_shift}
        G(\bar E) & = \frac{i \left( Z + \delta Z \right)}{\bar E + B} \left[1 - \frac{\delta B}{\bar E + B} + \cdots \right] \, .
    \end{align}
The energy shift $\delta B$ can now be identified as the coefficient of $-i Z/(\bar E + B)^2$ (see for instance Ref.~\cite{weinbergQuantumTheoryFields1995a}).

On the other hand, the two-point function can be expanded as
    \begin{align}
        G(\bar E) & = \frac{\Sigma_1}{1 + i C_0 \Sigma_1} + \frac{(-i C_2) (M_N \bar E) \Sigma_1^2}{\left( 1 + i C_0 \Sigma_1 \right)^2} + \frac{(-i C_4) (M_N \bar E)^2 \Sigma_1^2}{\left( 1 + i C_0 \Sigma_1 \right)^2} - \frac{C_2^2 (M_N \bar E)^2 \Sigma_1^3}{\left( 1 + i C_0 \Sigma_1 \right)^3} + \cdots \, ,
    \end{align}
where each coupling is a function of the subtraction velocity $\nu$.
In the sector of the theory with only potential terms, this is equivalent to time independent perturbation theory in ordinary quantum mechanics.
Thus, the shift in the binding energy can be obtained by inserting the \gls{lo} expression of the self-energy and manipulating the denominators of the second and third terms until reaching the form of Eq.~\eqref{eq:2pt_shift}.
The last term requires more care.
This term corresponds to a second-order perturbation theory calculation with the $C_2$ interaction, so we must explicitly separate the contribution of the deuteron in this term.
This is equivalent to calculating
    \begin{align}
        \bra{\psi} V^{(0)} \frac{1}{E - H} V^{(0)} \ket{\psi} - \frac{1}{E+B} \abs{ \bra{\psi} V^{(0)} \ket{\psi} }^2 \, ,
    \end{align}
where $H$ is the full \gls{lo} Hamiltonian containing $V^{(-1)}$.
All together the binding energy to \gls{n2lo} is
    \begin{align}
        B & = \frac{1}{M_N} \left( \frac{4 \pi}{M_N C_0} \right)^2 + \frac{1}{2 \pi} C_2 \left( \frac{4 \pi}{M_N C_0} \right)^5 + \frac{7}{16 \pi^2} M_N C_2^2 \left( \frac{4 \pi}{M_N C_0} \right)^8 - \frac{1}{2 \pi} C_4 \left( \frac{4 \pi}{M_N C_0} \right)^7
    \end{align}

We use the scattering length and effective range of the \gls{av18} potential \cite{wiringaAccurateNucleonnucleonPotential1995} without the electromagnetic interaction as the boundary value (i.e. at $\nu = m_{\pi}/M_N$) of the \gls{vRG} equations.
Electromagnetic corrections to the shape parameter $P$ are also expected to be small, so we use the Nijmegen value \cite{deswartLowEnergyNeutronProtonScattering1995a}.
In the deuteron channel, these are
    \begin{align}
         a_{np} & = 5.402 \ \text{fm} \, , \label{eq:scattering_parameter:a} \\
         r_{np} & = 1.752 \ \text{fm} \, , \label{eq:scattering_parameter:r} \\
         P_{np} & = 0.040 \ \text{fm}^{-3} \, . \label{eq:scattering_parameter:P}
    \end{align}
The LECs at $\nu = m_\pi/M_N$ are given in terms of these parameters according to
    \begin{align}
        C_0(m_\pi/M_N) & = \frac{4 \pi  a_{np}}{M_N} \, , \label{eq:normalization:c0} \\
        C_2(m_\pi/M_N) & = \frac{ 4 \pi}{M_N} \frac{  a_{np}^2  r_{np}}{2} \, , \label{eq:normalization:c2} \\
        C_4(m_\pi/M_N) & = \frac{4 \pi}{M_N}  a_{np}^3 \left( \frac{1}{4}  r_{np}^2 + \frac{ P_{np}}{ a_{np}}  \right) \, . \label{eq:normalization:c4}
    \end{align}
    
\begin{figure}[t]
    \centering
    \includegraphics[width=0.8\columnwidth]{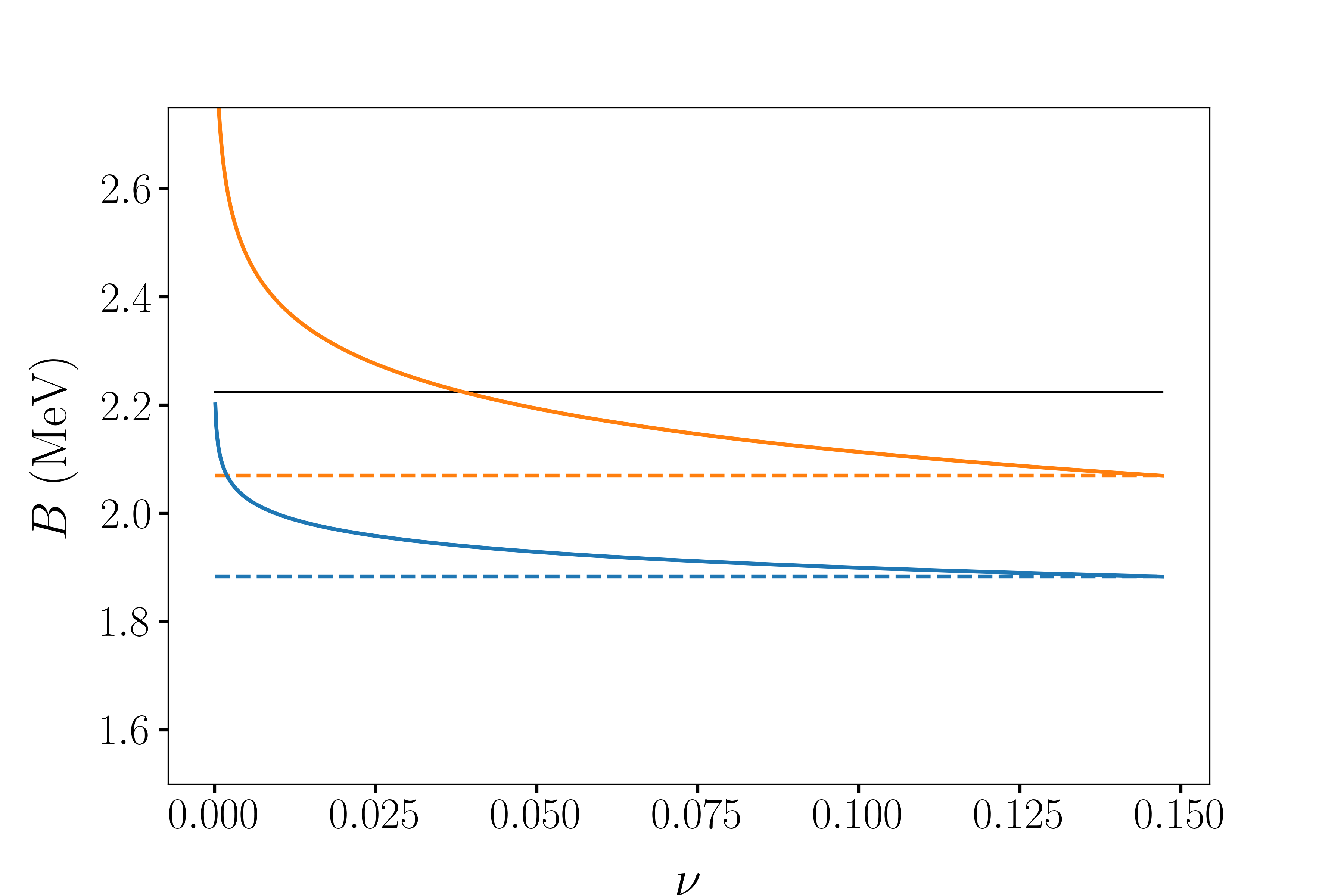}
    \caption{The deuteron binding energy as a function of the subtraction velocity. The solid black line is the experimental value. The blue (orange) dashed line is the fixed order \gls{nlo} (\gls{n2lo}) result while the blue (orange) solid line is the renormalization group improved \gls{nlo} (\gls{n2lo}) result. }
    \label{fig:deuteron_binding}
\end{figure}

The result for the deuteron binding energy at \gls{nlo} and \gls{n2lo} is shown in Fig.~\ref{fig:deuteron_binding}.
First, we can compare the values of the renormalization group improved binding energies at each order in the \gls{eft} to the values at the hard scale $\nu = m_\pi/M_N$.
When the subtraction velocity is $\nu \approx 0.04$ (corresponding to momenta around $38$ MeV), there is a shift in the binding energy of about $2.5\%$ at \gls{nlo}.
At \gls{n2lo}, the improvement shifts the binding energy by about $7\%$.
Moreover, the improvement at \gls{n2lo} causes the predicted binding energy to intersect the experimental value $B = 2.224575$ MeV around $\nu \approx 0.04$.
Second, we can compare the shift in the predicted binding energy at different orders in the \gls{eft} for the same subtraction velocities.
At the hard scale $\nu = m_\pi/M_N$, the binding energy at \gls{n2lo} is about $10\%$ larger than the result at \gls{nlo}.
At $\nu \approx 0.04$, the \gls{n2lo} is around $17\%$ larger than the \gls{nlo} result.

\section{Summary}
In this work, we have performed an analysis of the role of radiative corrections in the \NN system.
Using \gls{eft} techniques helps to organize the role of different strong and electromagnetic effects in a systematic expansion.
Additionally, we performed the first direct application of the \gls{vRG} in a nuclear \gls{eft}.
This allows us to sum the leading logarithm series into the potential coefficients.
We then provided evidence that the \gls{vRG} generates a percent level shift in the binding energy of the deuteron.
It is possible that similar corrections will play an important role in other light nuclei.
This prediction will be more robust when reliable \NN observables can be calculated in lattice QCD at the physical pion mass in order to match the couplings of this \gls{eft}.

The ultrasoft renormalization of the leading order potential in chiral \gls{eft} can be analyzed with similar techniques.
First, the one-pion exchange potential will be written as a four-fermion operator where the LEC is determined by the axial coupling $g_A$ and the pion decay constant $F_\pi$ at the breakdown scale of chiral \gls{eft} in the absence of electroweak effects. 
Then the tree-level potential will be dressed with an ultrasoft photon that leads to an anomalous dimension similar to Eq.~\eqref{eq:ultrasoft:c0}, only $C_0$ is replaced by $(g_A/F_\pi)^2$ up to a factor of 2.
Also, the contact potential proportional to $C_0$ will acquire a nonzero soft anomalous dimension driven by pion exchange.
Renormalizing the potential at higher orders will be significantly more difficult.

The running couplings obtained in this work can also be incorporated into other \gls{pionless} calculations or in \textit{ab initio} methods for nuclear physics that make use of \gls{pionless} potentials derived with dimensional regularization.
In this way, this renormalization group study can impact a variety of theoretical work relevant for ongoing experiments including $\beta$-decay, $\mu$-capture, and muonic atom spectroscopy.


\backmatter

\bmhead{Acknowledgements}

We would like to thank Sonia Bacca, Wouter Dekens, Aneesh Manohar, Matthias Schindler, and Roxanne Springer for interesting discussions.
This work was supported in part by the Deutsche Forschungsgemeinschaft (DFG) through the Cluster of Excellence ``Precision Physics, Fundamental Interactions, and Structure of Matter'' (PRISMA${}^+$ EXC 2118/1) funded by the DFG within the German Excellence Strategy (Project ID 390831469
).

\bibliography{radiative_references}

\end{document}